\documentclass[showpacs,pra,twocolumn,amsmath,amssymb]{revtex4}

\usepackage{amsthm}
\usepackage{amssymb}
\usepackage{graphicx}
\usepackage{color}
\usepackage{cancel}
\usepackage{soul}

\begin{document}

\title{Monogamy properties of quantum and classical correlations}

\author{Gian Luca Giorgi}

\affiliation{IFISC (UIB-CSIC),
Instituto de F\'isica Interdisciplinar y Sistemas Complejos, Campus Universitat de les Illes Balears,
E-07122 Palma de Mallorca, Spain}

\date{\today}
\begin{abstract}
In contrast with entanglement, as measured by concurrence, in general, quantum discord does not possess the property of monogamy, that is, 
there is no  tradeoff between the  quantum discord shared by a pair of subsystems and the quantum discord 
that both of them can share with a third party. Here, we show that, as far as monogamy is considered, quantum discord of pure states is equivalent to the entanglement of formation.
This result allows one to  analytically prove that none of the pure three-qubit states belonging to the 
subclass of W states is monogamous.
A suitable physical interpretation of the meaning of the correlation information as a quantifier of monogamy for the total information is also given.
Finally, we prove that, for rank 2 two-qubit states, discord and classical correlations are bounded from above by single-qubit von Neumann entropies.
\end{abstract}


\maketitle

\section{Introduction}

Entanglement, first recognized as the characteristic trait of quantum mechanics~\cite{epr}, 
has been used for a long time as the main indicator of the quantumness of correlations. Indeed, as shown in Ref.~\cite{linden}, for pure-state computation, exponential
speed-up occurs only if entanglement grows with the size of the system. 
However, the role played by entanglement in mixed-state computation is less clear. 
For instance, in the so-called deterministic quantum computation with one qubit
(DQC1)  protocol~\cite{DQC1}, quantum speed-up can be achieved using factorized states. As shown in Ref.~\cite{datta}, speed-up could be due to the presence of another quantifier, the 
so called quantum discord~\cite{zurek,henderson}, which is defined as the difference between two quantum analogs of the classical mutual information. 

The relationship between entanglement and quantum discord has not  been completely understood, since they seem to capture different properties of the states. 
 In Ref.~\cite{mdms}, it is 
shown that even if quantum discord and entanglement are equal for pure states, mixed states maximizing  discord 
in a given range of classical correlations are actually separable. The relation between discord and entanglement has been discussed in Refs.~\cite{cornelio,streltsov,piani}, and an operational meaning in terms of state merging has been proposed in~\cite{madhokcavalcanti}.

Recently, the use of quantum discord has been extended to multipartite states.
A measure of genuinely multipartite quantum correlations has been introduced in ~\cite{noi}.
In Ref.~\cite{sarandy},  an attempt to generalize the definition of quantum discord in multipartite systems based
on a collective measure has been proposed.
In Ref.~\cite{chakrabarty},
the authors proposed different generalizations of quantum discord depending on the measurement protocol performed.
Entanglement in multipartite systems has been shown to obey monogamy in the case of qubits~\cite{ckw,osborne} and continuous variables~\cite{adesso}.
Monogamy means that if two subsystems are highly correlated, the correlation between them and other parties is bounded. 
Prabhu \textit{et al.} proved that, unlike entanglement, quantum discord is in general not monogamous~\cite{prabhu}.
They also suggested, based on numerical results, that W states are likely to violate monogamy.

In this Brief Report, using the Koashi-Winter formula~\cite{winter}, we prove that, for pure states, quantum discord and entanglement of formation obey the same monogamy relationship, while, for mixed states,
distributed discord exceeds distributed entanglement.
Then, we give an analytical proof of the violation of monogamy by all W states. Furthermore, we suggest the use of the interaction information~\cite{libro} as a measure of monogamy
for the mutual information. Finally, as a further application of Koashi-Winter equality, we prove, the conjecture on upper bounds of quantum discord and classical correlations
formulated by Luo \textit{et al.}~\cite{luo} for rank 2 states of two qubits.

\section{Quantum discord}
In classical information theory, mutual information between parties $A$ and $B$ is defined as ${\cal J}(A:B)=H(A)-H(A|B)$, 
where $H(.)$ is the Shannon entropy and $H(A|B)$ is the conditional Shannon entropy of $A$ after $B$ has been measured.
An equivalent formulation [${\cal I}(A:B)=H(A)+H(B)-H(AB)$] can be obtained
 using Bayes' rules, because of which ${\cal I}(A:B)={\cal J}(A:B)$. On the other hand, if we try to quantize these quantities, 
replacing probabilities with density matrices and the Shannon entropy with the Von Neumann entropy, their counterparts differ substantially~\cite{zurek}.
The quantum mutual information is defined as
\begin{equation}
\mathcal{I}_{A,B}=S(\varrho_A)+S(\varrho_B)-S(\varrho_{A,B}),
\end{equation}
where $S(.)$ is the von Neumann entropy and $\varrho_{A(B)}$ are the reduced states after tracing out party $B(A)$, 
while the quantized version of  ${\cal J}(A:B)$ measures the  classical part of the correlations~\cite{henderson}  and it is given by 
\begin{equation}
{\cal{J}}_{A,B}=\max_{\{E_j^B\}}[S(\varrho_A)-S(A|\{E_j^B\})],\label{clas}
\end{equation}
 with the conditional entropy defined as $S(A|\{E_j^B\})=\sum_j p_j S(\varrho_{A|E_j^B})$, $p_j={\rm
Tr}_{AB}(E_j^B\varrho)$ and where $\varrho_{A|E_j^B}= E_j^B\varrho /{p_j} $ is the density
matrix after a positive operator valued measure (POVM) $\{E_j^B\}$ has been performed on $B$. In some cases, orthogonal measurements are enough to find the maximum in Eq.~(\ref{clas})~\cite{povm}.

 Quantum discord is thus defined as the difference between $\mathcal{I}$ and $\mathcal{J}$:
\begin{equation}
\label{eqdisc}
{\cal{D}}_{A,B}=\min_{\{E_j^B\}}\left[S(\varrho_B)-S(\varrho_{A,B})+S(A|\{E_j^B\})\right].
\end{equation}
Quantum discord can be considered as a measure of how much disturbance is caused when
trying to learn about party $A$ when measuring party $B$, and has been shown to be
null only for a set of states with measure zero~\cite{acin}.

Both classical correlations and quantum discord are asymmetric under the exchange of the two sub-parties (i.e.,
${\cal J}_{A,B}\neq{\cal J}_{B,A}$ and ${\cal D}_{A,B}\neq{\cal D}_{B,A}$).

While ${\cal J}$ is invariant under local unitary transformations and cannot increase under local operations and classical communication, ${\cal D}$ is not monotonic under local operations.
For instance, in~\cite{noisediscord}, it is shown how to create quantum correlations under the action of local  noise.

\section{Monogamy properties of quantum discord}
Given a measure of correlation $Q$, monogamy implies a tradeoff on bipartite correlations distributed along all the partitions $p_i$ ($i=1,\cdots N$):
\begin{equation}
 Q(p_1|p_2\cdots p_N)\ge\sum_{l\neq 1} Q(p_1|p_l).\label{monog}
\end{equation}
Coffman, Kundu, and Wootters~\cite{ckw} showed that this property applies to three-qubit states once the square of the concurrence~\cite{wootters} (${\cal C}^2$) plays the role of $Q$. 
The extension  of the proof to $n$-partite ($n>3$) qubit systems has been  given in Ref.~\cite{osborne}. As pointed out in~\cite{ckw}, 
however, entanglement of formation does not satisfy the criterion given in Eq.~(\ref{monog}). Then, even if people usually refers to entanglement as a monogamous quantity,
it would be worth paying attention to the entanglement monotone in use.

In trying to apply this property to quantum discord, Prabhu \textit{et al.} showed that monogamy is obeyed if and only if the interrogated interaction information
is less than or equal to the unmeasured interaction information~\cite{prabhu}. Then, the authors found through numerical simulations that the subset of W states are not monogamous, in contrast with Greenberger-Horne-Zeilinger (GHZ) states, which can be monogamous or not.

Here, we prove that, for pure states, the monogamy equations for quantum discord and for entanglement of formation coincide. 
Let us consider the pure tripartite state $|\psi_{ABC}\rangle$. Quantum discord of any of the couples of sub-parties is given by ${\cal D}_{i,k}=S(\varrho_k)-S(\varrho_l)+S(\varrho_{i|k})$,
where $S(\varrho_{i|k})=\min_{\{E_j^k\}}S(i|\{E_j^k\})$, while ${\cal D}_{i,jk}=S(\varrho_{i})$.
As shown in Ref.~\cite{winter}, the following relationships
between conditional entropies and entanglement of formation (${\cal E}$) holds: 
\begin{equation}
S(\varrho_{i|k})=S(\varrho_{l|k})={\cal E}(\varrho_{i,l}).\label{kw}
\end{equation}
This formula allows one to write 
\begin{equation}
 {\cal D}_{i,k}=S(\varrho_k)-S(\varrho_l)+{\cal E}(\varrho_{i,l}).\label{kwd}
\end{equation}
Using Eq.~(\ref{kw}), monogamy equation 
\begin{equation}
 {\cal D}_{A,B}+{\cal D}_{A,C}\le{\cal D}_{A,BC}\label{discmon}
\end{equation}
 is then equivalent to 
\begin{equation}
  {\cal E}_{A,B}+{\cal E}_{A,C}\le {\cal E}_{A,BC},\label{entmon}
\end{equation}
where $S(\varrho_{A})={\cal E}_{A,BC}$ has been employed.
The equality of conservation law for distributed entanglement of formation and quantum discord, even if not associated to monogamy, was already noticed by Fanchini {\it et al.}~\cite{fanchini}.
Because of this equivalence, the violation of Eq.~(\ref{entmon}) by W states, whose numerical evidence has been given in Ref.~\cite{prabhu}, admits an analytical proof.
Let us recall that, apart from local operations, a generic pure state of three qubits, belonging to the GHZ class, can be written as
$|\psi\rangle=\lambda_0 |0,0,0\rangle+\lambda_1 e^{i \theta} |1,0,0\rangle+\lambda_2 |1,0,1\rangle+\lambda_3 |1,1,0\rangle+\lambda_4 |1,1,1\rangle$~\cite{3qubit}.
The family of W states is obtained fixing $\lambda_4=0$. As shown by Coffman, Kundu, and Wootters~\cite{ckw}, W states  have zero three-tangle; that is, they obey
\begin{equation}
  {\cal C}^2_{A,B}+{\cal C}^2_{A,C}={\cal C}^2_{A,BC}.\label{concW}
\end{equation}
To show that Eq.~(\ref{concW}) implies $ {\cal E}_{A,B}+{\cal E}_{A,C}\ge {\cal E}_{A,BC}$, 
it is enough to note that ${\cal E}$ is a concave function of ${\cal C}^2$, since
${\cal E}=h[(1+\sqrt{1-{\cal C}^2})/2]$,
where $h$ is the binary entropy $h(x)=-x\log_2 x-(1-x)\log_2(1-x)$, and both ${\cal E}$ and ${\cal C}$ admit values between $0$ and $1$.
Then, if we apply the mapping from ${\cal C}^2$ to ${\cal E}$ to the three elements of Eq.~(\ref{concW}), we find  ${\cal E}_{A,B}+{\cal E}_{A,C}= {\cal E}_{A,BC}$
if ${\cal E}_{A,B}{\cal E}_{A,C}=0$ (i.e. for biseparable states) and ${\cal E}_{A,B}+{\cal E}_{A,C} > {\cal E}_{A,BC}$ otherwise.

\begin{figure}
    \includegraphics[width=8cm]{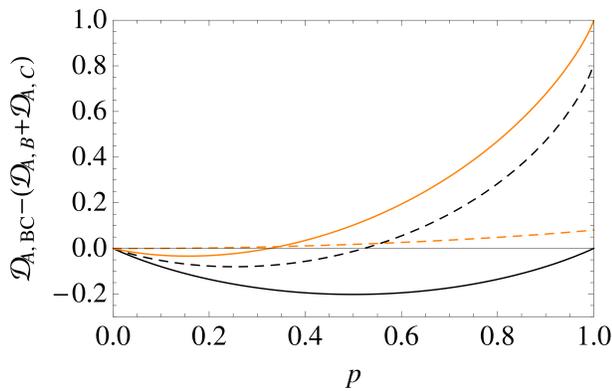}\\
  \caption{(Color online) Quantum discord monogamy of $|\tilde\psi(p,\epsilon)\rangle$, as quantified by ${\cal D}_{A,BC}-({\cal D}_{A,B}+{\cal D}_{A,C})= {\cal E}_{A,BC}-({\cal E}_{A,B}+{\cal E}_{A,C}) $,
as a function of $p$ for different values of $\epsilon$ (see the main text). 
States are monogamous where the respective curves are positive.
Black solid line is for $\epsilon=1$,  black dashed line is for $\epsilon=0.75$, 
 orange (gray) solid line is for $\epsilon=0.5$, and orange (gray) dashed line is for $\epsilon=10^{-2}$. According to the analytical proof, for W states ($\epsilon=1$)
${\cal E}_{A,B}+{\cal E}_{A,C}-S(\varrho_{A})$ is always positive (i.e., these states are never monogamous). For GHZ states, there exists a threshold value of $p$ above
 which monogamy is satisfied. This threshold goes to zero for vanishing $p$, since in this case qubit $A$ becomes factorized and all the related entanglement quantifiers vanish as well.}
\label{figure}
\end{figure}

As noticed in Ref.~\cite{prabhu}, GHZ states can be monogamous or not. Actually, a numerical analysis shows that about half of them do not respect monogamy.
To see a transition from observation to violation of monogamy, we consider the family of states
$|\tilde\psi(p,\epsilon)\rangle=\sqrt{p \epsilon} |0,0,0\rangle+\sqrt{p (1-\epsilon)} |1,1,1\rangle+\sqrt{(1-p)/2} (|1,0,1\rangle+ |1,1,0\rangle)$.
Note that $|\tilde\psi(1,1/2)\rangle$ is the maximally entangled GHZ state $(|0,0,0\rangle+ |1,1,1\rangle)/2$, while 
$|\tilde\psi(1/3,1)\rangle$ coincides with the maximally entangled W state $(|0,0,0\rangle+|1,0,1\rangle+ |1,1,0\rangle)/\sqrt{3}$.
For  $\epsilon=0$, qubit $A$ is factorized, and~(\ref{discmon}) becomes an equality.
In Fig.~\ref{figure}, ${\cal E}_{A,B}+{\cal E}_{A,C}-S(\varrho_{A})$ is plotted as a function of $p$ for different values of $\epsilon$.
As expected, for any $\epsilon\neq 1$, there is a threshold for $p$ above which the states are monogamous.

Once the assumption of pure state is relaxed, Eq.~(\ref{kw}) becomes an inequality: $S(\varrho_{i|k})\ge{\cal E}(\varrho_{i,l})$. Then,
${\cal D}_{A,B}+{\cal D}_{A,C}\ge S(\varrho_{B})-S(\varrho_{AB}) + { \cal E}_{A,C}+S(\varrho_{C})- S(\varrho_{AC}) +{\cal E}_{A,B}$, or, using the subadditivity of Von Neumann entropy,
\begin{equation}
 {\cal D}_{A,B}+{\cal D}_{A,C}\ge { \cal E}_{A,B}+{\cal E}_{A,C}.\label{discmix}
\end{equation}
Thus, for mixed states, monogamy of quantum discord has a stricter bound than monogamy of entanglement.

\section{Monogamy properties of classical and total correlations}

The search for monogamy of correlations can be extended to  ${\cal{J}}(\varrho)$. It is actually easy to note that, for pure tripartite states,
monogamies of quantum discord and classical correlations are complementary, that is
${\cal D}_{A,B}+{\cal D}_{A,C}-{\cal D}_{A,BC}=-({\cal J}_{A,B}+{\cal J}_{A,C}-{\cal J}_{A,BC})$.
To prove it, it is sufficient to observe that mutual information obeys
\begin{equation}
 {\cal I}_{A,BC}={\cal I}_{A,B}+{\cal I}_{A,C}.\label{minfmon}
\end{equation}
The generalization of Eq.~(\ref{minfmon}) to mixed states presents some interesting aspects. We have 
\begin{equation}
 {\cal I}_{A,BC}={\cal I}_{A,B}+{\cal I}_{A,C}+{\cal I}_{ABC}.\label{minfxmon}
\end{equation}
where $ {\cal I}_{ABC}$ is called correlation information~\cite{libro}.
  In the language of density matrices, it can be defined as
\begin{equation}
 {\cal I}_{12\cdots n}=\sum_{\vec{j}}(-1)^{||\vec{j}||+N+1} S(\varrho_{\vec{j}}),\label{intinf}
 \end{equation} 
 where $\vec{j}=\{j_1\dots j_r\}$ are all the possible strings containing integer numbers between $1$ and $n$, with $j_i\neq j_l$ for any $i\neq l$, and $||\vec{j}||=r$ counts the length of each string.
For instance, for a bipartite system, ${\cal I}_{AB}=S(\varrho_A)+S(\varrho_B)-S(\varrho_{AB})$ coincides with the ordinary mutual information, and in the tripartite case 
${\cal I}_{ABC}=-S(\varrho_A)-S(\varrho_B)-S(\varrho_C)+S(\varrho_{AB})+S(\varrho_{AC})+S(\varrho_{BC})-S(\varrho_{ABC})$. 
It can be checked that, for $n$ odd, ${\cal I}_{12\cdots n}=0$ for any pure state.

In classical information theory, the interaction information has been introduced with the aim of 
measuring the information that is contained in a given set of variables and that cannot be accounted for 
 considering any possible subset of them. It should then measure genuine $n$-partite correlations.
Actually,  ${\cal I}_{12\cdots n}$ can be negative. Thus, according to the criteria given, 
for instance, in Ref.~\cite{henderson}, it cannot be used as a correlation measure. Then, its meaning is widely debated. 
Equation~(\ref{minfxmon}) suggests that it plays the role played by the tangle  in the  distribution of ${\cal C}^2$~\cite{ckw},
since it is invariant under index permutation, and it can be called a ``mutual information tangle.''
When ${\cal I}_{ABC}$ is negative, it quantifies the lack of monogamy of the mutual information. As shown by Prabhu \textit{et al.},
monogamy of discord relies on the relationship between ${\cal I}_{ABC}$ and its interrogated version~\cite{prabhu}.

\section{Upper bound of quantum and classical correlations}

In Ref.~\cite{luo}, it has been conjectured that, given a bipartite state $\varrho_{AB}$, 
defined in the Hilbert space ${\cal H}_{A}\otimes {\cal H}_{B}$, the following upper bounds for quantum discord and classical correlations could exist:
\begin{eqnarray}
 {\cal D}_{A,B}&\le& \min [S(\varrho_{A}),S(\varrho_{B})],
\\{\cal J}_{A,B}&\le& \min [S(\varrho_{A}),S(\varrho_{B})].
\end{eqnarray}
It is trivial to prove the existence  of such an upper bound for entanglement of formation, its definition being based on the convex roof construction.
If inequality~(\ref{discmix}) were an equality, it would be easy to extend the proof to discord.
Actually, inequality~(\ref{discmix}) is telling us that distributed discord could exceed distributed entanglement, and these upper bounds could be violated.
While a a partial proof of the conjecture has been given in Ref.~\cite{zhang} using the language of quantum operations, a full proof for the case of rank 2 states of two qubits can be given using the Koashi-Winter formula. 
By applying a purification procedure,
we add an ancillary Hilbert space ${\cal H}_{C}$ and write a pure tripartite state $|\phi_{ABC}\rangle$ such that ${\rm Tr}_C |\phi_{ABC}\rangle\langle\phi_{ABC}|=\varrho_{AB}$. Since $\varrho_{AB}$ has rank 2, $|\phi_{ABC}\rangle$ is a three-qubit state.
As a consequence of Eq.~(\ref{kw}), we have 
\begin{eqnarray}
 {\cal D}_{A,B}&=& S(\varrho_B)-{\cal J}_{C,B},\\
{\cal J}_{A,B}&=& S(\varrho_{A})-{\cal E}_{A,C}.
\end{eqnarray}
Then, inequalities ${\cal D}_{A,B}\le S(\varrho_{B})$ and ${\cal J}_{A,B}\le S(\varrho_{A})$ are immediately verified.

Let us now separately discuss the cases  $S(\varrho_{A})>S(\varrho_{B})$ and  $S(\varrho_{A})<S(\varrho_{B})$.
In the first case, we only need to prove ${\cal J}_{A,B}\le S(\varrho_{B})$.
In Ref.~\cite{noi}, using the invariance under index permutation of the three tangle introduced by Coffman, Kundu, and  Wootters ~\cite{ckw}, we proved that, for the case of three qubits, if  $S(\varrho_{A})>S(\varrho_{B})$, then $S(\varrho_{A})+{\cal E}_{B,C}<S(\varrho_{B})+{\cal E}_{A,C}$.
This chain rule implies 
$S(\varrho_{A})-{\cal E}_{A,C}<S(\varrho_{B})-{\cal E}_{B,C}$, and then
\begin{equation}
 {\cal J}_{A,B}< S(\varrho_{B})-{\cal E}_{B,C}< S(\varrho_{B}),
\end{equation}
as we wanted to prove.

Assuming now $S(\varrho_{A})<S(\varrho_{B})$, we are left to show that ${\cal D}_{A,B} \le S(\varrho_A)$.
Writing explicitly ${\cal D}_{A,B}= S(\varrho_B)+{\cal E}_{A,C}-S(\varrho_C)$, we use the chain rule to write
$S(\varrho_{B})+{\cal E}_{A,C}<S(\varrho_{A})+{\cal E}_{B,C}$ and to obtain
\begin{equation}
 {\cal D}_{A,B} <S(\varrho_{A})-{\cal J}_{C,B}<S(\varrho_{A}).
\end{equation}
This ends the proof.

\section{Conclusions}
We have studied the monogamy properties of pure tripartite state.
We have shown that quantum discord and entanglement of formation obey the same monogamy relationship.
Applying this equivalence to the case of three qubits,  we have shown, by analytical demonstration, that, for all the  W states,
quantum discord is not monogamous, in contrast with GHZ states, where discord can be monogamous or not. 
In an example, we have shown the transition from monogamy to absence of monogamy for a subfamily of GHZ states.

The equivalence between quantum discord and entanglement of formation concerning monogamy raises a subtle question that it is worth considering.
While people usually claim that, for qubits, entanglement is monogamous, all we know is that there exists an entanglement monotone (the square of the concurrence) that is in fact monogamous.
By analogy, we can say that the results of Ref.~\cite{prabhu} do not exhaust the search for monogamy of quantum discord and other correlations, where monogamous  monotone indicators could be found.

Using the connection between discord end entanglement of formation, 
we have also shown that in the case of rank 2 states of two qubits, as conjectured by Luo {\it et al.}~\cite{luo}, quantum discord and classical correlations are bounded from above by the single-qubit Von Neumann entropies. A full proof cannot be given because, for mixed states, the equality of conservation law for distributed entanglement of formation and quantum discord is broken.

\acknowledgments{It is a pleasure to thank R. Zambrini and F. Galve for discussions and criticism. Funding from the CoQuSys (200450E566) project is acknowledged. The author is supported by the Spanish Ministry of Science and Innovation 
through the program ``Juan de la Cierva''.}


\begin{thebibliography}{10}


\bibitem{epr} A. Einstein, B. Podolsky, and N. Rosen, Phys. Rev. {\bf 47}, 777 (1935).

\bibitem{linden} R. Jozsa and N. Linden, Proc. R. Soc. A  {\bf 459}, 2011  (2003).

\bibitem{DQC1} E. Knill and R. Laflamme, Phys. Rev. Lett. {\bf 81}, 5672 (1998);
B. P. Lanyon, M. Barbieri, M. P. Almeida, and A. G. White, {\it ibid.}  {\bf 101}, 200501 (2008).


\bibitem{datta} 
A. Datta, A. Shaji, and C. M. Caves, Phys. Rev. Lett.  {\bf 100}, 050502 (2008).

\bibitem{zurek}  H.  Ollivier and W.  H.  Zurek,  Phys. Rev. Lett.  {\bf 88}, 017901 (2001).


\bibitem{henderson} L.  Henderson and V.  Vedral,  J. Phys. A  {\bf 34}, 6899 (2001).

\bibitem{mdms}  F. Galve, G. L. Giorgi, and R. Zambrini, Phys. Rev. A {\bf 83}, 012102 (2011).

\bibitem{cornelio} M. F. Cornelio, M. C. deOliveira, and F. F. Fanchini, Phys. Rev. Lett.  {\bf 107}, 020502 (2011). 

\bibitem{streltsov} A. Streltsov, H. Kampermann, and D. Bruss, Phys. Rev. Lett. 106, 160401 (2011).


\bibitem{piani}  M. Piani, S. Gharibian, G. Adesso, J. Calsamiglia, P. Horodecki, and A. Winter, Phys. Rev. Lett. {\bf 106}, 220403 (2011). 

\bibitem{madhokcavalcanti} V. Madhok and A. Datta, Phys. Rev. A {\bf 83}, 032323 (2011);   
D. Cavalcanti, L. Aolita, S. Boixo, K. Modi, M. Piani, and A. Winter, Phys. Rev. A  {\bf 83}, 032324 (2011).


\bibitem{noi}  G. L. Giorgi, B. Bellomo, F. Galve, and R. Zambrini, Phys. Rev. Lett. {\bf 107}, 190501 (2011).

\bibitem{sarandy}  C. C. Rulli and M. S. Sarandy, Phys Rev. A  {\bf 84}, 042109 (2011).

\bibitem{chakrabarty}  I. Chakrabarty, P. Agrawal, and A. K. Pati,  arXiv:1006.5784.


\bibitem{ckw} V. Coffman, J. Kundu, and W. K. Wootters, Phys. Rev. A {\bf 61}, 052306 (2000).

\bibitem{osborne} T. J. Osborne and F. Verstraete, Phys. Rev. Lett. {\bf 96}, 220503 (2006).


\bibitem{adesso} G. Adesso and F. Illuminati, New J. Phys. {\bf 8},  15 (2006);
G. Adesso, A. Serafini, and F. Illuminati, Phys. Rev. A {\bf 73}, 032345 (2006);
 T. Hiroshima, G. Adesso, and F. Illuminati, Phys. Rev. Lett. {\bf 98}, 050503 (2007). 


\bibitem{prabhu} R. Prabhu, A. K. Pati, A. Sen (De), and U. Sen,  arXiv:1108.5168. 


\bibitem{winter} M. Koashi and A. Winter,  Phys. Rev. A  {\bf 69}, 022309 (2004).

\bibitem{libro} T. M. Cover and J. A. Thomas, {\it Elements of Information Theory} (Wiley, New Jersey, 2006).

\bibitem{luo} S. Luo, S. Fu, and N. Li, Phys. Rev. A {\bf 82}, 052122 (2010).

\bibitem{povm} F. Galve, G. L. Giorgi, and R. Zambrini, EPL {\bf 96}, 40005 (2011).

\bibitem{acin} A. Ferraro, L. Aolita, D. Cavalcanti, F. M. Cucchietti, and
A. Acin, Phys. Rev. A {\bf 81}, 052318 (2010).

\bibitem{noisediscord} S. Campbell, T. J. G. Apollaro, C. Di Franco, L. Banchi, A. Cuccoli, R. Vaia, F. Plastina, and M. Paternostro, arXiv:1105.5548;
F. Ciccarello and V. Giovannetti, arXiv:1105.5551; A. Streltsov, H. Kampermann, and D. Bruss, Phys.
Rev. Lett. {\bf 107}, 170502 (2011).



\bibitem{wootters}  W. K. Wootters, Phys. Rev. Lett. {\bf 80}, 2245 (1998).



\bibitem{fanchini} F. F. Fanchini, M. F. Cornelio, M. C. deOliveira, and A. O.
Caldeira,  Phys. Rev. A {\bf 84}, 012313 (2011).


\bibitem{3qubit} A. Ac\'in, A. Andrianov, L. Costa, E. Jan\'e, J. I. Latorre, and R. Tarrach, Phys. Rev. Lett. {\bf 85}, 1560 (2000);
W. D\"ur, G. Vidal, and J. I. Cirac, Phys. Rev. A  {\bf 62}, 062314 (2000);
A. Ac\'in, D. Bruss, M. Lewenstein, and A. Sanpera, Phys. Rev. Lett. {\bf 87}, 040401 (2001).


\bibitem{zhang} L. Zhang and J. Wu, arXiv:1105.2993.



\end{thebibliography}
\end{document}